# AI Sensing and Intervention in Higher Education: Student Perceptions of Learning Impacts, Affective Responses, and Ethical Priorities

Bingyi Han
School of Computing and Information Systems
University of Melbourne
Melbourne, Australia
Department of Computer Science
Saarland University
saarbrücken, Germany
bingyi.han@unimelb.edu.au

Ying Ma
School of Computing and Information Systems
University of Melbourne
Melbourne, Australia
yima3@student.unimelb.edu.au

Simon Coghlan
Computing and Information Systems
University of Melbourne
Melbourne, Australia
simon.coghlan@unimelb.edu.au

Dana McKay
School of Computing Technologies
RMIT University
Melbourne, Australia
dana.mckay@rmit.edu.au

George Buchanan
School of Computing Technologies
RMIT University
Melbourne, Australia
g.r.buchanan@gmail.com

Wally Smith
School of Computing & Information Systems
University of Melbourne
Melbourne, Australia
wsmith@unimelb.edu.au

## Abstract

AI technologies that sense student attention and emotions to enable more personalised teaching interventions are increasingly promoted, but raise pressing questions about student learning, well-being, and ethics. In particular, students' perspectives about AI sensing-intervention in learning are often overlooked. We conducted an online mixed-method experiment with Australian university students (N=132), presenting video scenarios varying by whether sensing was used (in-use vs. not-in-use), sensing modality (gaze-based attention detection vs. facial-based emotion detection), and intervention (by digital device vs. teacher). Participants also completed pairwise ranking tasks to prioritise six core ethical concerns. Findings revealed that students valued targeted intervention but responded negatively to AI monitoring, regardless of sensing methods. Students preferred system-generated hints over teacher-initiated assistance, citing learning agency and social embarrassment concerns. Students' ethical considerations prioritised autonomy and privacy, followed by transparency, accuracy, fairness, and learning beneficence. We advocate designing customisable, social-sensitive, non-intrusive systems that preserve student control, agency, and well-being.

## 1 Introduction

AI-enabled sensing and intervention systems are increasingly promoted in higher education [103] to support data-informed teaching [61, 82] and personalised learning [4]. These technologies often use camera-based inputs, such as eye movement and facial cues, to monitor students' attention and emotional engagement [12, 75]. Such systems can then either generate automated intervention suggestions for teachers (e.g., [6, 75, 92]), or provide real-time tailored feedback to students (e.g., through Intelligent Tutoring Systems [77]). These systems aim to enhance teachers' situational awareness [63, 91, 139] and deliver timely, individualised learning support to students.

Despite rapid technical advances, research on AI in Education (AIEd) often overlooks deeper student-centred and learning factors [17, 22]. Prior work has largely taken a technology-centred perspective to understand which design features can improve system effectiveness, efficiency, and usability (e.g., [84]). Yet technical performance alone does not guarantee positive or meaningful educational experiences (e.g., [54]). Particularly underexplored are students' perspectives on how various system features might influence their affective experiences [78] and learning [27]. These aspects can significantly shape students' engagement with AIEd [16, 47, 53]. For instance, discomfort from being observed by AIEd may trigger behavioural change [53] and undermine the technology's effectiveness [54]. Misalignment with student needs and expectations could precipitate disengagement or resistance, even if the tools are technically effective.

Crucial AI ethics considerations are equally underexamined in AIEd [60, 103]. For instance, improving personalisation and accuracy requires extensive data, threatening privacy [129]. Timely intervention can aid learning efficiency but override self-directed learning [55, 89], diminishing autonomy. Improved AI accuracy often increases model opacity, while greater transparency can reduce system performance [9]. Such ethical trade-offs present design dilemmas. Furthermore, designing AI without appreciating how students perceive value tensions risks building technically sound but socially unacceptable systems. Student-centric AIEd thus requires consideration of relevant ethical concerns from the outset [43].

To address these gaps, we conducted a two-stage experimental study with 132 Australian university students via an online survey. Stage 1 examined student preferences for AI sensing-intervention features. Participants viewed six video scenarios featuring screen mockups of AI systems, which varied along three dimensions: 1) whether sensing was used; 2) sensing approach (gaze-tracking attention detection vs. facial-based emotion detection); 3) form of intervention (system-generated instruction vs. teacher-mediated assistance). Participants rated their responses using five-point Likert items after each scenario and provided open-ended feedback. Stage 2 explored how students weigh ethical value trade-offs, asking participants to rank concerns through pairwise comparison tasks. Our study answers two research questions:

- **RQ1** *(Stage 1): How do students perceive AI sensing and intervention systems with various features?*
- **RQ1 (a)**: *How do students evaluate the likely learning impacts of these systems?*
- **RQ1 (b)**: *How do students affectively respond to these systems?*
- **RQ2** *(Stage 2): How do students prioritise ethical concerns regarding AI sensing and intervention systems?*

Our findings reveal that students' affective responses and beliefs about learning and ethics indicate a consistently negative reception to the monitoring element of such technology (regardless of whether gaze-based or facial-based detection was involved) in some of its current forms. In contrast, students showed a clear preference for system-generated over teacher-initiated assistance, citing desires for agency and concerns about embarrassment before peers. Regarding ethical views, students were most concerned about autonomy and privacy, suggesting they highly value control over learning processes and data in interactions with educational technology. Our study makes three contributions to HCI and AIEd design. It:

- Foregrounds human-centred dimensions in student-AI interaction, examining crucial affective and pedagogical responses that are rarely evaluated. Our findings challenge prevailing notions of effectiveness metrics, highlighting the importance of incorporating emotional well-being, social dynamics, and ethical values into AIEd design and assessment.

- Presents new empirical evidence for how students prioritise ethical values when engaging with AIEd. This offers a foundation for aligning AI tools with student values and provides a transferable method for ethical design in other sensitive domains where value tension arises.
- Proposes five student-centred design implications for AIEd: minimise intrusive sensing; enable student-triggered, customisable, and lightweight system interventions; redesign teacher-mediated interventions to preserve student agency and control; implement dynamic consent mechanisms and transparency to reinforce trust; and provide private feedback channels before teacher involvement.

## 2 Related work

This section reflects the two-stage structure of this study. Section 2.1 reviews previous work for Stage 1, justifying the hypotheses to answer RQ1. Section 2.2 reviews ethical literature for Stage 2, introducing ethical principles examined to address RQ2.

### 2.1 Student experiences with AI sensing-intervention

*2.1.1 AI sensing-intervention.* Educational AI increasingly emphasises real-time sensing of learner states to enable adaptive feedback and enhance instructors' situational awareness [2, 34, 141]. Timely data-informed learning support has been recognised as an effective mechanism for education [1, 21]. While sensor-equipped or instrumented classrooms have explored embedding computing into physical infrastructures (e.g., pressure-sensitive chairs or mouses [5], wearable devices with biometric sensors that physiologically identify arousal states [10, 51, 138]), these approaches pose logistical or financial challenges [1]. These practical limitations have generated significant interest in visual approaches in AI sensing [1], using computer vision and machine learning to capture students' visual and behavioural cues (e.g., gaze direction [2, 64, 85], facial expressions [12, 42], body gestures [74, 90], head orientation [1, 104, 105]) and infer cognitive attention, emotional engagement, and interactions [34, 131, 141].

***AI Sensing:*** Among camera-based sensing modalities, gaze-based attention detection and facial emotion detection are predominant [1, 3]. Using existing cameras, such as those in student laptops, makes AI sensing practical for widespread adoption.

*Gaze-based Attention Detection* leverages eye-tracking to infer moment-to-moment attentional state [64] (e.g., focus vs. distraction [3, 69]). It has been applied to assess student performance in complex tasks like surgical simulations [75, 111] and to mitigate mind-wandering with timely interventions [33, 64].

*Facial Emotion Detection* uses facial landmarks with machine learning trained to classify emotional expressions [12, 42, 115] (e.g., excitement [74], confusion, and frustration [19, 68]). This potentially offers rich situational information on student engagement [133] and real-time learning experiences, informing tools such as teacher-facing dashboards (e.g., EmoDash [42]).

***AI Intervention:*** Insights from AI sensing of attention and emotion can be directly used to trigger instructional interventions. Two main forms exist:

*Teacher-Mediated Assistance*: Teacher-facing systems [61, 82] can provide insight about individual students or class-wide progress, such as who is struggling or disengaged [123], enabling teachers to adjust instruction [6, 7, 42, 48], target their attention [62] and offer timely support [31, 62, 64, 136]. This positions teachers as "orchestrators of when and how to use AIEd tools" [88].

*System-Generated Instruction*: Student-facing systems [14] include Intelligent Tutoring Systems (ITS) [115], which automatically deliver hints, feedback, or re-engagement messages [77] directly to students. For instance, attention-aware ITS such as GazeTutor [33] redirect attention when students become distracted, and Guru [64] and ATS [84] adapt feedback and provide hints based on emotional cues.

While attracting much interest, these promising technologies still lack strong empirical evidence and robust pedagogical grounding [128]. Furthermore, little is known about how users, especially students, respond to these approaches. Addressing this gap is essential for student-centred AIEd design.

*2.1.2 Gaps in understanding students' experiences with AIEd.* Understanding students' perceptions of AIEd is critical for aligning technology with users' values and needs [8, 18], and for achieving intended pedagogical outcomes. When students feel uncomfortable being monitored by AIEd, they may alter their behaviours, pandering to or even deceiving the system [53], which can reduce AIEd accuracy and effectiveness and disrupt students' sense of agency and trust [54, 56]. Thus, beyond technical performance of AIEd, we must examine how students experience, interpret, and respond to these technologies.

While some empirical studies evaluate students' experiences with specific AIEd technologies such as ITS and affective-aware systems, they largely emphasise student performance metrics like pre-/post-learning gains according to scores [80, 84], faster task-completion times [97], usability ratings [80, 84, 97], and technology acceptance [27]. Such findings demonstrate statistically significant learning improvements and system effectiveness [46], but risk narrowing student attitude and acceptance to system efficiency alone. Deeper human aspects of learning experiences, such as students' learning status and autonomy, motivation, high-order thinking, and self-efficacy, are rarely examined [27]. Learning is more than task completion; it is a complex phenomenon that involves developing curiosity, agency, and personal growth. Without deeper investigation into how students experience or perceive AIEd, we risk relying on evaluations that are not pedagogically meaningful or user-centred [8, 18].

Two recent speculative studies surface students' nuanced reactions to classroom AI observation [53] and personalised intervention [56]. This research offers early insights into emotional and pedagogical concerns about AIEd, pointing to heightened anxiety and disrupted learning [55, 89] and pedagogical interactions [53, 86, 135]. Although they also note positive expectations (e.g., timely support, data-informed educational insights), these studies do not isolate which specific features trigger students' negative emotional or learning-related responses. Nor do they address how students weigh trade-offs between negative emotional costs of being sensed and perceived learning benefits expected from such data-driven intervention. This leaves a critical gap in understanding how combinations of sensing and intervention features shape students' attitudes.

Hence, this study addresses the lack of research on students' learning-related beliefs and affective responses to AI sensing-intervention through RQ1. RQ1(a) – *How do students evaluate the likely learning impacts of these systems?* – builds on prior work highlighting students' expectations for personalisation, efficiency [50, 117], academic improvement [80, 84], and learning autonomy [53, 56]. RQ1(b) – *How do students affectively respond to these systems?* – draws on evidence that students' emotional responses significantly shape their attitudes toward AI [70] and their learning engagement [54], including enjoyment [53], confidence [81, 84, 97], anxiety [53], discomfort [53], and fear of mistakes [53].

*2.1.3 Hypotheses in Stage 1.* To address RQ1, we generated a set of hypotheses on students' perceptions of different AI sensing-intervention approaches. Given that such systems typically involve a sequential process that first collects students' behavioural data through sensing and delivers adaptive support through intervention, we break down the evaluation to examine perceptions of each component individually.

As noted, research on ITS has shown statistically significant improvement in learning outcomes (e.g., pre/post-test score gains and reduced task-completion time [80, 84, 97]), while other studies highlight students' positive expectations of more efficient and personalised learning [50, 117]. However, studies have also reported negative student perceptions toward AI monitoring, including emotional discomfort and a sense of disrupted learning [53, 56]. Building on these contrasting findings, we propose:

- **H1** - Students perceive AI systems that rely on sensing more negatively than systems that do not use sensing – both in terms of (a) potential learning impacts and (b) affective responses.

It also seems reasonable to expect that students may perceive gaze-based attention detection more positively than facial-based emotion detection, due to differences in how personal the collected data feels and also the perceived pedagogical relevance. Facial emotion recognition, which relies on sensitive biometric data (e.g., facial expressions), is legally classified as high-risk in school settings under the GDPR and EU AI Act [38], and has received much greater ethical and regulatory scrutiny than gaze-based sensing. In contrast, gaze-based attention detection typically involves observable behavioural cues (where the student is looking) and may feel less intrusive. Students may therefore feel more comfortable with gaze-based systems. Pedagogically, attention is a well-established indicator of learning outcomes [107], whereas the connection between AI-detected emotions and learning is more ambiguous and context-dependent. Accordingly, *we propose:*

- **H2** - Students perceive gaze-based attention detection more positively than facial-based emotion detection – both in terms of (a) potential learning impacts and (b) affective responses.

Some research suggests some students prefer automated assistance over human support to avoid feeling judged [54, 113], whereas other studies report a preference for human-delivered help, citing perceived empathy [39, 40]. This tension becomes particularly relevant in the context of AI sensing-intervention systems, where automated support might feel more private but less human. Meanwhile, ITS's offering personalised automated hints and explanations have been shown to improve learning outcomes and receive positive evaluation from students [76, 130]. To explore this further, we specifically examine which form of assistance students feel more comfortable receiving. We propose:

- **H3** - Students perceive system-generated assistance more positively than teacher-mediated assistance – both in terms of (a) potential learning impacts and (b) affective responses.

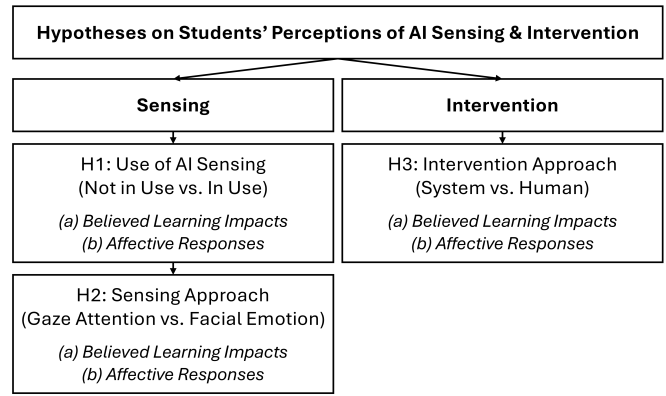

Figure 1: Overview of study hypotheses and experimental factors (addressing RQ1).

In summary (see Fig. 1), H1 relates to whether AI sensing should be used at all. H2 involves considering whether the type of sensing matters (gaze-based vs. facial-based). H3 relates to asking how, regardless of sensing, the intervention should be delivered (system-generated vs. teacher-mediated assistance). All hypotheses align with our overarching research aim of understanding how students evaluate AI sensing-intervention tools.

## 2.2 Students' ethical priorities for AI sensing-intervention.

Ethical values are central to human-centred computing [101]. Recent review studies [124], academic critics [59, 60], and policy frameworks (e.g., UNESCO [95], World Economic Forum [137]) highlight key ethical concerns for AIEd. These concerns primarily involve data privacy, bias and fairness, transparency, autonomy, accuracy, and impacts on pedagogy [60]. These values can conflict in real-world implementation. For instance, improving model accuracy requires big data, thereby compromising privacy [129]. Enhancing algorithm performance can reduce transparency [9], as it can involve using complex 'black box' models that are difficult to interpret or explain [106]. Moreover, opting out of AI-enabled education to preserve autonomy may lead students to forgo potential learning benefits, raising fairness concerns. Students themselves report experiencing such trade-offs, such as that between obtaining the benefits of participating in AI-enabled systems and having their privacy compromised by them [109]. Such trade-offs mean that fully meeting all ethical principles simultaneously can be costly or impossible. In practice, developers often make decisions about which values to prioritise or compromise [72]. Importantly, compromising any ethical principle, even with good intent, may harm students.

No existing study has examined how students themselves evaluate these trade-offs. While prior research has documented students' general attitude to AI, often highlighting privacy and fairness as the highest-rated concerns [65], these insights remain high-level. Yet users' attitudes toward ethical risk are highly context-dependent [121]. For instance, people accept algorithmic decision-making in social media recommendations and reject it in financial services [65]. Explainability may be subordinated to accountability, fairness, security, privacy, and accuracy in tax fraud detection [71]. In education, overlooking how students navigate value tensions could inadvertently undermine agency and trust. Hence, we pose *RQ2: How do students prioritise ethical concerns regarding AI sensing and intervention systems?* Below, we explain key ethical concepts examined in Stage 2.

*2.2.1 Ethical constructs examined in Stage 2.* To ground this study's ethical prioritisation task in Stage 2, we draw on widely recognised frameworks. Jobin et al. [66] identified five core AI ethics principles: non-maleficence, privacy, justice and fairness, transparency, and responsibility. Influenced by bioethics, Floridi and Cowls [44] stressed beneficence, non-maleficence, justice, autonomy, and explicability. Such broad principles, however, need contextualisation within educational settings [59, 60]. Holmes et al.'s [60] influential AIEd-specific framework highlighted privacy, data security, fairness, bias, autonomy and agency, transparency, accountability, and inclusion. Six recurring ethical values from these frameworks form the basis of our Stage 2 ranking task. Here, we explain these ethical concepts and their significance for AI sensing-intervention.

*(Learning) Beneficence:* Beneficence refers to assisting and benefitting others [44]. In AI ethics, beneficence demands that AI systems be capable of achieving benefit [96]. In the AIEd context, beneficence requires that technologies meaningfully support students' learning, engagement, and educational development [103]. Importantly, beneficence in AIEd must be grounded in evidence of actual educational value [128].

*Privacy:* Privacy is a core expression of the non-maleficence principle [45] of "do no harm" [96] and requires safeguarding individuals' personal information. AI systems may collect, store, and process large volumes of sensitive personal data [95]. Privacy concerns include preventing data misuse, hacking, or leakage [81] and respecting individuals' control over how their information is gathered and shared. Persistent privacy worries have been reported by students [58, 67, 79, 99, 125, 129], educators [50, 83], and school officials [24], particularly regarding excessive data collection [95] and breaches of personal information [81].

*Accuracy:* Accuracy, another aspect of non-maleficence, requires that AI systems avoid harm [45] through correct interpretations and decisions [96]. In AIEd, inaccurate interpretation and predictions concerning behaviour can mislabel students and trigger disruptive interventions [26, 49, 54]. For instance, remote exam AI proctoring tools may falsely flag innocent behaviours, causing distress and unfair allegations of cheating [28]. Accuracy failures in AIEd have consequences for fairness, well-being, and trust (e.g., [28]).

*Autonomy:* Autonomy refers to students' ability to pursue meaningful goals [60] and make informed decisions [45]. Respecting autonomy generally involves the right to consent to AI use, opt out, exercise control over data [58], and retain some level of agency in learning [56]. The degree of choice that students have concerning AI in education depends on circumstances, but a perceived lack of choice can negatively affect student attitudes to AIEd [56].

*Fairness:* Fairness, very frequently mentioned in discussions of AIEd [124], requires that AI models avoid or mitigate discriminatory or unjust consequences [52, 121] to ensure equitable treatment across diverse backgrounds [93]. This is particularly pressing in AIEd applications such as personalised content delivery, performance evaluation and prediction, and facial emotional recognition (e.g., [37]). Here, biased training data or model assumptions [87, 94] can cause unfair student outcomes [13, 127, 140] regarding race, ethnicity, nationality, and gender [60]. For instance, automated essay-grading systems may assign lower scores to essays by Black over White students for equivalent content [26, 49]. Notably, fairness has been shown to most strongly influence users' satisfaction with algorithms, followed by transparency [119].

*Transparency:* Transparency in AI systems can build trust [118] and empower users [25]. It can involve explanations [112] to stakeholders about data collection, algorithmic function [100], and system limitations, and is linked to informed consent [93] and respect for autonomy [28]. Low transparency in AIEd can leave students uncertain about what is being measured or how interventions are triggered [53], generating feelings of surveillance and anxiety and producing self-censorship [53]. Transparency can be a requirement [28] for student understanding, trust, and meaningful participation [53].

## 3 Study Method

This study employed a two-stage design. Stage 1 (RQ1) examined students' preferences for AI sensing-intervention features through an online experiment with video scenarios, followed by Likert-scale responses and open-ended reports. Stage 2 (RQ2) examined ethical priorities where participants rated and ranked ethical principles. Ethical approval for the study was granted by the University of Melbourne.

### 3.1 Stage 1: Scenario-Based Experiment

*3.1.1 Materials and variable design.* We employed a within-subjects experimental design with three independent variables: whether sensing was used (sensing-in-use vs not-in-use), sensing modality (gaze-based attention detection vs facial-based emotion-detection, only when sensing was in-use), and intervention form (system-generated vs. teacher-mediated intervention).

- Use of AI sensing: whether the scenario depicted a system that relied on AI sensing (*sensing-in-use*) or a system that did not use sensing (*not-in-use*). In the experiment, this was conveyed by informing participants that AI sensing was operating (*sensing-in-use*) or by omitting such information (*not-in-use*). In practice, students' learning status could also be inferred through non-sensing approaches (e.g., time spent on tasks or interaction patterns within the software), though we did not detail these alternative approaches in the scenarios.
- Sensing approach: whether the sensing involved *gaze-based attention detection* or *facial-based emotion-detection*.
- Intervention form: whether the learning intervention was offered via a *system-generated hint* (displayed on the screen) or via a *human teacher* (meaning the issuing of a screen alert that a human teacher would soon visit the student to assist).

**Experimental Scenarios.** To represent these combinations (see Table.1), we created six video scenarios using simple screen-based prototypes (Appendix A) and presented them within a Qualtrics survey. Each video simulated a student's view of a learning task within a typical learning management system, overlaid with animated features representing AI sensing-intervention functionalities. These visual materials were accompanied by text instructions and voice-overs within the videos to provide participants with context for the prototypes they were viewing. This commentary emphasised that our goal was not to criticise the technology but rather to explore implications to inform better design. All instructions were carefully worded with neutral terms such as "... the system observes you", avoiding expressions that might bias participants' perceptions.

*3.1.2 Procedure.* Participants entered the study via the Qualtrics survey platform. After reading an overview introducing that the research focused on exploring their attitudes to intelligent technologies in education, they provided consent for participation. We informed them that two sequential stages of data collection followed.

In stage 1, participants watched six scenarios (Table.1) and were asked to imagine themselves using the systems displayed. After each scenario, participants responded to a set of five-point Likert

Table 1: Scenario presentation sequence in Stage 1 [a]

**Sensing not-in-use (S1, S2 randomised)**
S1: System-generated hint (System hint)  |  S2: Human-mediated assistance (Human assistance)
**Sensing in-use (S3-S6 randomised)**
S3: Gaze-based attention detection + System hint  |  S4: Gaze-based attention detection + Human assistance
S5: Facial-based emotion detection + System hint  |  S6: Facial-based emotion detection + Human Assistance

[a] For the first two scenarios (S1 or S2, presented in randomised order), participants were not informed that sensing was involved. For the remaining trials (S3-S6, represented in randomised order), participants were explicitly informed that AI-based sensing was operating. These scenarios varied by sensing approach and intervention form.

Table 2: Likert items presented after each scenario presentation

| Constructs and Likert Statements | Related Concepts |
|---|---|
| *Belief about learning [Cronbach's Alpha = 0.948, high reliability]* | |
| personalisation [50, 117] - 'This system would be useful in meeting my individual learning needs.' | Perceived usefulness [35] |
| efficiency [50, 117] - 'Using this system could make my learning process easier.' | Job-fit [126] |
| control of pace [53, 56] - 'I would have complete control of my own learning at the pace I want when using this system.' | Perceived behavioural control [71] |
| distraction [53] - 'It would distract me to think that my learning progress was observed.' | |
| grade performance [80, 84] - 'Using this system could help me gain better grades on the learning tasks.' | Social-cognitive theory (individual esteem, sense of accomplishment) [30] |
| over-helping [56] - 'Using this system would give me so much help that I don't really learn.' | |
| *Affective response [Cronbach's Alpha = 0.957, high reliability]* | |
| enjoyable [53] - 'My learning experience with this new feature is more enjoyable.' | Attitude toward behaviour [36, 142] |
| confident [81, 84, 97] - 'Using this system helps me feel more confident in my learning.' | Emotional reaction to technology (ERT) model [16] |
| anxious [53] - 'I feel anxious about using the system to learn.' | Positive affect negative affect schedule (PANAS scale) [132] |
| uncomfortable [53] - 'It is uncomfortable to think that my learning progress is observed.' | Students' emotional responses to observational AI [53] |
| worried [53] - 'I am worried about the system flagging that I am struggling.' | |
| fearing mistakes [53] - 'I fear making mistakes, as it is observing my learning progress.' | |

items to record their beliefs about the likely learning impact of the technologies and their affective responses. Upon completing all 6 scenarios, participants were given an open-ended question that invited them to enter free text to describe in richer detail their responses in relation to all the systems viewed. This step was included to gain qualitative insights to interpret patterns in the Likert ratings, and for the researchers to learn about any aspects of experience that had been overlooked.

*3.1.3 Measures and scale validation.* We designed a set of Likert-scale items to assess participants' beliefs about learning and affective responses to sensing and intervention technologies. Existing technology acceptance scales (e.g., [73]) were not suitable to capture the specific affective and pedagogical concerns raised in prior observational and qualitative studies of AIEd systems (e.g., [53, 56]). Therefore, we devised new items informed by the literature reviewed in Section 2.1.2 (first column of Table.2) and grounded in relevant theoretical frameworks on technology perception and user experience. Each item was mapped to an established psychological concept or cognitive theory (second column of Table.2) to support construct validity. The items were piloted with 10 students to ensure clarity and ease of interpretation, with minor wording revisions made based on their feedback.

Internal consistency of the scales was assessed using Cronbach's alpha [32]. Both constructs of *Beliefs about Learning* and *Affective Responses* showed high reliability. Each construct combined six items with both positive and negative items. We applied reverse-scoring to the negative items to align their direction with the positive items. Cronbach's Alpha was then computed, resulting in reliability scores of 0.948 for *Beliefs about Learning*, and 0.957 for *Affective Responses*, both well above the recommended threshold of 0.70 [102]. These reliability scores indicate that the items consistently measured students' perceptions of learning impacts and affective responses, which we used to test whether these two dimensions were viewed positively.

*3.1.4 Analysis.* For Likert-scale quantitative responses, we conducted non-parametric Mann–Whitney U tests with Bonferroni corrections to compare conditions across items. For open-ended responses, we employed qualitative content analysis with descriptive coding [116]. The first author read all responses, developed an initial descriptive code list capturing manifest content (e.g., surveillance discomfort, peer embarrassment), and coded the full dataset. Given the brevity and clarity of responses and to maintain coding consistency, a single researcher conducted the coding. To ensure analysis transparency, weekly peer-debrief sessions were conducted within the research team to review code definitions, category boundaries, and disconfirming cases. The codebook was iteratively refined through constant comparison, collapsing overlapping codes into higher-order categories that summarised the most salient concerns and preferences. For example, codes including *"forced teacher help"*, *"cannot choose when to get help"*, and *"loss of control"* were collapsed into *"Disrupted learning experience, learning autonomy, and control"*. We report category frequencies (number of participants mentioning a category) to convey salience across short responses (typically 1–3 sentences), without inferring statistical significance.

Table 3: Statements of potential ethical concerns

| Statement Tag | Statement |
| --- | --- |
| Privacy_1 | Protecting my personal data from breaches. |
| Privacy_2 | Sharing information about my activities with others, including teachers. |
| Privacy_3 | Sharing information about my emotional states with others, including teachers |
| Autonomy_1 | Having the fact that I am interacting with AI hidden from me. |
| Autonomy_2 | Not being asked for my consent before using the system. |
| Autonomy_3 | Not being given the option to opt out of using the system. |
| Fairness_1 | Not getting the same amount of support as other students get. |
| Fairness_2 | The system discriminating against students based on, e.g., their race and gender. |
| Fairness_3 | The system only being available to privileged students. |
| Accuracy_1 | The system inaccurately interpreting my behaviour and emotions. |
| Accuracy_2 | The system failing to recognise when I need help. |
| Accuracy_3 | The system offering too much help when I don't need it. |
| Transparency_1 | Not really understanding how the system's AI algorithm works. |
| Transparency_2 | Not being informed about how all the data collected about me is being used. |
| Transparency_3 | Not being informed about the potential risks and flaws of the system. |
| Learning_1 | Not having access to this system and therefore waiting for help for too long. |
| Learning_2 | Not having access to this system, and therefore not getting enough attention from teaching staff. |
| Learning_3 | The system not respecting my individual learning pace and preferences. |

Table 4: Ethical constructs and their representative statements of concern, for paired comparison ranking

| Ethical Concept | Presented Statement |
| --- | --- |
| Privacy: | Students not controlling their private information held by the technology. |
| Autonomy: | Students not having choices about whether and how this technology is used. |
| Fairness: | Students not being treated equally and fairly by the technology. |
| Accuracy: | The technology not accurately determining when students need help. |
| Transparency: | Students not knowing how the technology evaluates their attention and emotions. |
| Learning Beneficence: | The technology not helping students to learn effectively. |

### 3.2 Stage 2: Ethics Pairwise Ranking Tasks

After completing the scenario-based evaluations in Stage 1, participants proceeded to Stage 2, which involved two sequential tasks.

*3.2.1 Design and procedure.* First, participants rated their level of concern about 18 statements describing potential ethical issues, using a five-point Likert scale (1=not at all concerned, 5=extremely concerned). These statements were grouped under six broad ethical concepts: Privacy, Autonomy, Fairness, Accuracy, Transparency, and Learning Beneficence (Table.3). This task aimed to provide a descriptive overview of participants' ethical concerns depicted across all the viewed scenarios in Stage 1.

Next, participants completed 15 pairwise comparisons between representative statements from six core AI ethics concepts: Privacy, Autonomy, Fairness, Accuracy, Transparency, and Learning Beneficence (Table.4). In each comparison, participants were asked to select which of the two presented statements they were more concerned about. Each item in Table 4 was a single overarching statement designed to broadly represent the core ethical concern it belonged to. The order of both rating items and pairings was randomised for each participant.

Finally, participants completed a brief demographic questionnaire reporting their gender, age range, major and degree of study, and self-evaluation of general AI knowledge.

*3.2.2 Item justification.* The items and statements used in Stage 2 reflect our review of salient ethical principles from established frameworks (Section 2.2.1). In the rating task in Table 3, each principle was represented by three specific statements. Internal consistency of the scales in Table 3 was assessed using Cronbach's alpha [32], demonstrating high and acceptable internal consistency for *Privacy* (0.90), *Autonomy* (0.82), *Transparency* (0.84), and *Fairness* (0.72), which exceeded the recommended threshold of 0.70 [102]. *Accuracy* (0.68) and *Learning* (0.67) showed reliability levels close to the 0.70 threshold. For *Learning*, it is likely because the three items pull in different directions: Learning_1 and Learning_2 focus on risks of *not using* AI, while Learning_3 concerns risks *when AI is used*. We therefore present the Learning dimension result at the item level rather than relying on mean scores [102].

### 3.2.3 Analysis.
Participants' concern levels across 18 rated items (Table.3) were presented using descriptive statistics. Pairwise comparisons (Table.4) were analysed with the Bradley-Terry Model [20] to estimate the probability of each concern being prioritised over another.

## 3.3 Participants

We recruited 141 participants via the crowdsourcing platform Prolific, with 9 excluded based on an attention check to ensure data quality. A power analysis suggested a minimum of 106 participants were required, based on a medium effect size [29] f=0.35 with a significance level of 0.05 at 80%. Participants were all university students in Australia, with English as their first language and a minimum approval rate of 95% on the Prolific platform. We did not screen participants for AI literacy, as the aim was to reflect the real-world diversity of AI understanding within higher education settings (however, they were asked to self-evaluate their general AI knowledge in the demographic questionnaire at the end). In our sample, 50% identified as male, 47% identified as female, and 3% identified as non-binary or third/gender. All participants were compensated for their time and contribution with a payment of GBP £13.64 per hour, matching the local minimum wage in Australia.

## 4 Findings

Our final dataset included 132 participants aged 18~24 (51%), 25~35 (44%), 45~54 (5%), with none below 18 or above 55. All participants were current university students, including undergraduates (67%), postgraduates (17%), PhD (9%), and others such as honours and diplomas (7%). Majors taken were arts and humanities (10%), business (8%), education (4%), engineering (11%), computer science and IT (16%), health sciences (20%), natural sciences (5%), social sciences (15%), and other (11%). In terms of AI literacy, most participants reported basic (43%) or intermediate knowledge (43%), with fewer indicating advanced (11%), no knowledge (2%), or expert knowledge (1%).

## 4.1 Stage 1: Believed learning impacts and affective response towards AI sensing-intervention

We first present students' quantitative ratings from Stage 1 to address RQ1: *How do students perceive AI sensing and intervention systems with various features?* Hypothesis testing was based on two constructs: Beliefs about Learning and Affective Responses, each consisting of multiple items. For construct scores, items with negative valence (e.g., "*distraction*") were reverse-coded so that higher values consistently reflect more positive perceptions. Item-level scores were reported in their original direction to provide nuanced insights into specific perceptions.

### 4.1.1 Sensing-in-use vs. not-in-use (Hypothesis 1).
Table 5 shows Likert ratings when sensing was not-in-use (scenarios 1-2) versus sensing-in-use (scenarios 3-6).

In the sensing not-in-use condition, participants rated their beliefs on learning impacts relatively positively. They agreed that the intervention system would support *personalisation*, *efficiency* and *grade performance*, while holding neutral views on having *self-control of their learning pace*, being *distracted*, and receiving *too much help* from the system. Their affective responses were also generally neutral.

In the sensing-in-use condition, ratings shifted sharply negative on both believed learning impacts and affective responses. Participants disagreed that the system would support *personalisation*, *efficiency*, and *self-control of learning pace*. They deemed it *distracting*, finding the experience *unenjoyable* and not *confidence-boosting*, while anticipating feeling *anxious*, *uncomfortable*, *worried about being flagged as struggling*, and *afraid of making mistakes*.

Five of six learning beliefs items (personalisation, efficiency, control of pace, grade performance, distraction) and all six items from affective responses (enjoyable, confident, anxious, uncomfortable, worried, fearing mistakes) were significantly less positive when sensing was disclosed, with only over-helping unaffected. For example, perceptions on improved efficiency declined from Mean=3.60 (agree) in sensing not-in-use scenarios to Mean=2.56 (disagree) in sensing-in-use scenarios. Disclosure of sensing led to negative responses on every item apart from *over-helping*. These results support H1 - Students perceive AI systems that rely on sensing more negatively than systems that do not use sensing, both in terms of (a) potential learning impacts and (b) affective responses.

### 4.1.2 Gaze-based attention detection vs. Facial-based emotion detection (Hypothesis 2).
Table 6 compares sensing approaches across scenarios 3–6 when students were informed that sensing was in operation.

Across both conditions, participants reported relatively negative perceptions in both learning beliefs and affective dimensions. No significant differences were found between gaze-based attention detection and facial-based emotion detection on any of the items. Thus, we found no support for *H2 - Students perceive gaze-based attention detection more positively than facial-based emotion detection, both in terms of (a) potential learning impacts and (b) affective responses.*

### 4.1.3 System-generated vs. Human-mediated assistance (Hypothesis 3).
Table 7 compares intervention approaches (human assistance vs. system-generated hints) when students were not informed that sensing was in operation (scenarios 1-2). Table 8 compares intervention approaches when sensing was in use (scenarios 3-6).

In the sensing not-in-use condition (Table.7), system-generated hints were perceived significantly more positively than human teacher assistance for 3 items of believed learning impacts (*efficiency, control of peace, distraction*), with no difference for 3 items (*personalisation, grade performance, over-helping*). System-generated hints were also rated significantly more positively in terms of all the items of affective response. In the sensing-in-use condition (Table.8), system-generated hints again were rated significantly more positively than human teacher assistance, across nearly all items in believed learning impacts except *over-helping*, and all items in affective responses. These results support H3 - *Students perceive system-generated assistance more positively than human-delivered assistance, both in terms of (a) potential learning impacts and (b) affective responses.*

A notable perception shift was observed when comparing intervention ratings across the sensing not-in-use condition (Table.7) and sensing-in-use condition (Table.8). As shown, perceptions of both intervention approaches worsened substantially once sensing

Table 5: Effect of Sensing Use on Beliefs about Learning and Affective Responses [a]

|  | Sensing Not-In-Use | | Sensing In-Use | | Significance | | |
| --- | --- | --- | --- | --- | --- | --- | --- |
|  | Mean | SD | Mean | SD | U-statistic | P-value | Adjusted P-value |
| **Belief about Learning** | **3.28** | **0.89** | **2.51** | **0.89** | 105309.50 | 0.00 | 0.00 |
| personalisation | 3.59 | 1.12 | 2.63 | 1.30 | 99703.50 | 0.00 | 0.00 |
| efficiency | 3.60 | 1.15 | 2.56 | 1.30 | 101653.00 | 0.00 | 0.00 |
| control of pace | 3.04 | 1.36 | 2.22 | 1.23 | 94830.00 | 0.00 | 0.00 |
| grade performance | 3.50 | 1.03 | 2.66 | 1.18 | 98763.50 | 0.00 | 0.00 |
| distraction | 2.97 | 1.17 | 3.90 | 1.13 | 39976.50 | 0.00 | 0.00 |
| over-helping | 3.08 | 1.20 | 3.12 | 1.16 | 69395.00 | 0.65 | 1.00 |
| **Affective Responses** | **2.85** | **1.07** | **1.96** | **0.95** | 36202.50 | 0.00 | 0.00 |
| enjoyable | 3.07 | 1.21 | 2.00 | 1.13 | 104535.50 | 0.00 | 0.00 |
| confident | 2.77 | 1.09 | 2.09 | 1.08 | 95773.00 | 0.00 | 0.00 |
| anxious | 3.03 | 1.41 | 4.22 | 1.13 | 36894.50 | 0.00 | 0.00 |
| uncomfortable | 3.26 | 1.39 | 4.33 | 1.06 | 38535.00 | 0.00 | 0.00 |
| worried | 2.70 | 1.25 | 4.00 | 1.18 | 32420.50 | 0.00 | 0.00 |
| fearing mistakes | 3.01 | 1.34 | 3.82 | 1.17 | 46517.00 | 0.00 | 0.00 |

[a] The response score is based on the Five-point Likert scale (1=Strongly disagree, 5=Strongly agree). Construct scores (shaded rows) were calculated by reverse-coding negatively worded items (e.g., 'distraction') so that higher values indicate more positive perceptions. Item-level scores are presented in their original direction.

Table 6: Effect of Sensing Approach on Beliefs about Learning and Affective Responses [a]

|  | Gaze Attention | | Facial Emotion | | Significance | | |
| --- | --- | --- | --- | --- | --- | --- | --- |
|  | Mean | SD | Mean | SD | U-statistic | P-value | Adjusted P-value |
| **Belief about Learning** | **2.50** | **0.95** | **2.52** | **0.93** | 34814.40 | 0.75 | 1.00 |
| personalisation | 2.57 | 1.29 | 2.70 | 1.30 | 33503.50 | 0.27 | 1.00 |
| efficiency | 2.49 | 1.29 | 2.62 | 1.30 | 33385.00 | 0.25 | 1.00 |
| control of pace | 2.17 | 1.22 | 2.27 | 1.23 | 33643.00 | 0.31 | 1.00 |
| grade performance | 2.67 | 1.21 | 2.66 | 1.15 | 35443.50 | 0.97 | 1.00 |
| distraction | 3.85 | 1.13 | 3.94 | 1.12 | 33481.00 | 0.26 | 1.00 |
| over-helping | 3.05 | 1.17 | 3.20 | 1.15 | 32660.50 | 0.11 | 1.00 |
| **Affective Responses** | **1.94** | **0.93** | **1.97** | **0.96** | 34791.00 | 0.74 | 1.00 |
| enjoyable | 1.95 | 1.09 | 2.05 | 1.17 | 34034.00 | 0.42 | 1.00 |
| confident | 2.07 | 1.07 | 2.11 | 1.09 | 34726.50 | 0.70 | 1.00 |
| anxious | 4.24 | 1.11 | 4.19 | 1.16 | 36181.50 | 0.61 | 1.00 |
| uncomfortable | 4.34 | 1.06 | 4.31 | 1.07 | 35970.00 | 0.70 | 1.00 |
| worried | 4.00 | 1.18 | 4.00 | 1.17 | 35370.00 | 1.00 | 1.00 |
| fearing mistakes | 3.81 | 1.15 | 3.83 | 1.20 | 34583.00 | 0.64 | 1.00 |

[a] The response score is based on the Five-point Likert scale (1=Strongly disagree, 5=Strongly agree). Construct scores (shaded rows) were calculated by reverse-coding negatively worded items (e.g., 'distraction') so that higher values indicate more positive perceptions. Item-level scores are presented in their original direction.

Table 7: Effect of Intervention Form (Sensing Not-In-Use Scenario) on Beliefs about Learning and Affective Responses [a]

| Sensing Not-In-Use Scenario | Human Assist | | System Hint | | Significance | | |
|---|---|---|---|---|---|---|---|
| | Mean | SD | Mean | SD | U-statistic | P-value | Adjusted P-value |
| **Belief about Learning** | **3.09** | 0.92 | **3.47** | 0.82 | 6513.00 | 0.00 | 0.00 |
| personalisation | **3.41** | 1.24 | **3.77** | 0.97 | 7549.50 | 0.02 | 0.29 |
| efficiency | **3.29** | 1.25 | **3.91** | 0.95 | 6377.50 | 0.00 | 0.00 |
| control of pace | **2.48** | 1.29 | **3.59** | 1.19 | 4831.00 | 0.00 | 0.00 |
| grade performance | **3.44** | 1.19 | **3.56** | 0.96 | 8366.00 | 0.42 | 1.00 |
| distraction | **3.17** | 1.19 | **2.77** | 1.13 | 10604.00 | 0.00 | 0.05 |
| over-helping | **2.94** | 1.21 | **3.23** | 1.19 | 7734.50 | 0.07 | 0.83 |
| **Affective Responses** | **2.26** | 0.95 | **3.47** | 0.94 | 3104.00 | 0.00 | 0.00 |
| enjoyable | **2.62** | 2.00 | **3.52** | 1.06 | 5153.00 | 0.00 | 0.00 |
| confident | **2.45** | 1.00 | **3.10** | 1.09 | 5868.50 | 0.00 | 0.00 |
| anxious | **3.84** | 1.26 | **2.22** | 1.03 | 14538.50 | 0.00 | 0.00 |
| uncomfortable | **3.96** | 1.18 | **2.56** | 1.23 | 13897.00 | 0.00 | 0.00 |
| worried | **4.05** | 1.09 | **2.84** | 1.26 | 13457.50 | 0.00 | 0.00 |
| fearing mistakes | **3.68** | 1.22 | **2.33** | 1.08 | 13891.50 | 0.00 | 0.00 |

Table 8: Effect of Intervention Form (Sensing-In-Use Scenario) on Beliefs about Learning and Affective Responses [a]

| Sensing-In-Use Scenario | Human Assist | | System Hint | | Significance | | |
|---|---|---|---|---|---|---|---|
| | Mean | SD | Mean | SD | U-statistic | P-value | Adjusted P-value |
| **Belief about Learning** | **2.29** | 0.86 | **2.73** | 0.97 | 25527.50 | 0.00 | 0.00 |
| personalisation | **2.36** | 1.24 | **2.90** | 1.30 | 27214.50 | 0.00 | 0.00 |
| efficiency | **2.23** | 1.20 | **2.88** | 1.31 | 25629.00 | 0.00 | 0.00 |
| control of pace | **1.86** | 1.04 | **2.58** | 1.29 | 24052.50 | 0.00 | 0.00 |
| grade performance | **2.49** | 1.18 | **2.84** | 1.15 | 29229.00 | 0.01 | 0.00 |
| distraction | **4.05** | 1.08 | **3.74** | 1.15 | 41088.00 | 0.01 | 0.00 |
| over-helping | **3.17** | 1.16 | **3.08** | 1.17 | 36966.50 | 0.36 | 1.00 |
| **Affective Responses** | **1.70** | 0.88 | **2.21** | 0.95 | 21834.50 | 0.00 | 0.00 |
| enjoyable | **1.73** | 1.02 | **2.27** | 1.17 | 25475.50 | 0.00 | 0.00 |
| confident | **1.89** | 1.00 | **2.30** | 1.12 | 27571.00 | 0.00 | 0.00 |
| anxious | **4.50** | 1.00 | **3.93** | 1.189 | 46439.50 | 0.00 | 0.00 |
| uncomfortable | **4.55** | 0.93 | **4.11** | 1.138 | 44194.50 | 0.00 | 0.00 |
| worried | **4.28** | 1.08 | **3.72** | 1.206 | 45604.00 | 0.0 | 0.00 |
| fearing mistakes | **4.10** | 1.11 | **3.54** | 1.169 | 45789.00 | 0.00 | 0.00 |

[a] The response score is based on the Five-point Likert scale (1=Strongly disagree, 5=Strongly agree). Construct scores (shaded rows) were calculated by reverse-coding negatively worded items (e.g., 'distraction') so that higher values indicate more positive perceptions. Item-level scores are presented in their original direction.

was disclosed, across all items in learning impacts and affective responses (Appendix B presents a side-by-side comparison from synthesising Tables 7 and 8 to illustrate this shift).

When sensing was not-in-use, participants viewed system-generated hints relatively positively. They tended to agree that these hints supported *personalisation* (mean=3.77) and *efficiency* (mean=3.91). Regarding affective responses, they disagreed that such intervention would cause *anxiety* (mean=2.22), *discomfort* (mean=2.56) and *fear of making mistakes* (mean=2.33). However, once informed that sensing underpinned these interventions, evaluations worsened substantially on all items. System hints were now seen as more *distracting* (mean=3.74, changed from 2.77) and emotionally taxing, with sharp increases in *anxiety* (mean=3.93, changed from 2.22). *discomfort* (mean=4.11, changed from 2.56), and *fear of mistakes* (mean=3.54, changed from 2.33).

A similar pattern held for human teacher assistance. Initially rated neutral-to-negative when not using sensing, these ratings became even more negative when linked to sensing, with high levels of *anxiety* (mean=4.50) and *discomfort* (mean=4.55). Together, these results show that while students may accept intervention features on their own, once tied to sensing, they triggered strongly negative reactions, undermining both perceived learning benefits and emotional comfort.

## 4.2 Stage 1: Qualitative reports about learning impacts and affective responses to AI sensing-intervention

We next present qualitative findings from participants' open-ended responses, which provide insights that help explain the quantitative preferences in Section 4.1.

*4.2.1 Anxiety and discomfort about constant monitoring (N=78).* The most expressed concern was discomfort at constant observation by AI. Participants frequently reported feelings of "anxious" (P50), "uncomfortable" (P39), and "distracted" (P90). One called it "an invasion of personal space and autonomy" (P24). Participants voiced strong reactions to both sensing methods and their intrusiveness:

> "That is horrible and dystopian, having a system constantly track and analyse my physical attributes would be extremely anxiety-inducing and stop my learning altogether. If this happened in my classes, I would start coming to class wearing masks and sunglasses to stop it, or I would not come to class at all" (P41)

*4.2.2 Disrupted learning experience, learning autonomy, and control (N=66).* Participants felt that being monitored would distract them from learning. Constant sensing could cause pressure to constantly perform well or appear focused. With either attention or emotion detection, participants described feeling hyper-conscious of their appearance rather than concentrating on learning.

> "I have to act and force myself to look more attentive, which is more distracting." (P90)

> "Tracking eye gaze and emotions ... pressured me to be fully focused on my work for the entire class, which sometimes is just not possible." (P32)

Participants were dissatisfied with potential disruptions to the learning process from teacher assistance triggered automatically by the AI system. Many felt uncomfortable with additional social interactions with teachers when they were "not ready to be helped" (P128). They preferred self-control and autonomy over when and how assistance was provided, rather than passively receiving interventions initiated by a teacher.

> "Having a computer assume I need help is a little patronising. It could be beneficial to suggest a hint, but the computer calling over a teacher for me is not okay... University students are adults and should be respected as such. I want to choose when I feel that I need help and ask for it myself." (P122)

> "I like the system where it lets me (not the teacher) know that I am struggling and gives me hints. It would be even better if it asks me whether I need the hint or not, and then I can decide when to use the hint or whether to use the hint at all. Calling the teacher without asking me is distracting and feels forced." (P62)

Perceived disruptions to learning autonomy often added stress to learning:

> "Alerting a teacher without my consent would make me feel very anxious as I usually try to get the task done alone" (P50).

> "Having my teacher automatically notified every time I struggle or lose focus would be very distracting and anxiety-inducing." (P91)

*4.2.3 Social dynamics and implications: peer perception, judgment, and embarrassment (N=25).* Another key concern was being identified as struggling or flagged for needing help in front of others, particularly teachers. Participants feared unwanted attention, embarrassment, and peer judgment. One said: "If everybody could see the teacher coming to help me whenever AI detected I was struggling, that has a certain social impact." (P68). Another worried it would "make it easier for peers to identify who's struggling and cause more bullying" (P93). Another said:

> "I feel having the teacher notified whenever the AI thinks that I'm struggling is detrimental to the class and to my learning experience. Having the teacher attend to a particular student over and over again would draw the teacher's attention away from other students and could cause the student receiving ongoing assistance embarrassment or discomfort." (P36)

Students preferred keeping their emotional and learning status private. One suggested how the system could be improved:

> "If it gathered classroom data on average completion time and offered a hint without anyone knowing, that would be the best method." (P31)

*4.2.4 System inclusiveness and accommodating diverse learner needs (N=32).* Participants questioned whether the system could accommodate diverse learning styles, disabilities, cultural nuances, and neurodivergence. They worried it lacked flexibility and could misinterpret behaviours, disadvantaging those not conforming to expected patterns.

> "I am on the autism spectrum—my learning style would trigger unnecessary attention. I often display unusual facial expressions and don't make normal eye contact." (P25)

> "It's not inclusive. It will make some students, like those with ADHD, very embarrassed and more anxious because their behaviour might be misinterpreted by the AI." (P60)

> "Facial recognition may be designed from a white Anglo-Saxon perspective." (P71)

Participants also worried about the system's inability to account for everyday emotional fluctuations, leading to misinterpretation of their emotional state and inappropriate intervention.

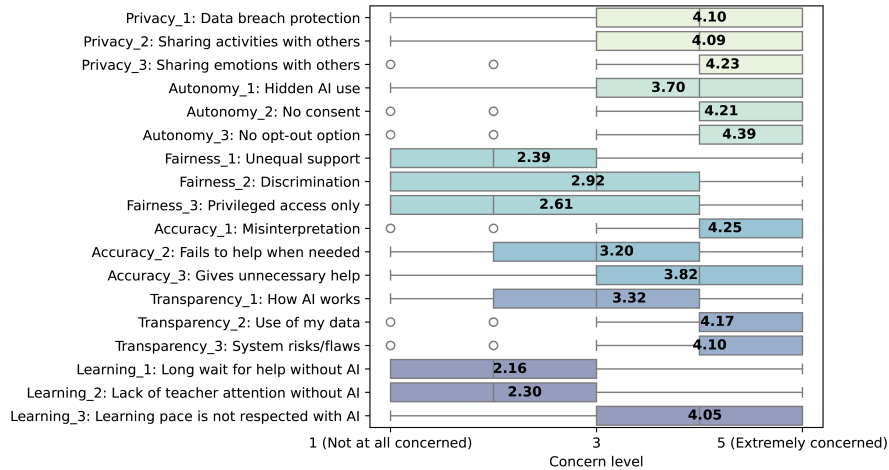

Figure 2: Mean ratings for statements of ethical concerns (see Table.3 for full item wording). The response score is based on the Five-point Likert scale rating from 1 (Not at all concerned) to 5 (Extremely Concerned). Bars positioned further to the right indicate stronger concern.

"We shouldn't be shamed for having a bad day. This system pressures us to be constantly alert and emotionally stable, which is unrealistic." (P55)

"My facial expressions don't always show what I feel...if I am very focused, I could look upset/angry, especially to a computer." (P122)

These qualitative findings help explain students' negative reactions to sensing (H1) regardless of the two approaches (H2), which they found intrusive and disruptive to learning. Social dynamics surrounding teacher intervention explained students' preference for system-generated hints (H3), which allowed them to retain control and avoid social consequences such as peer judgment and embarrassment at being publicly flagged for struggling. Overall, these themes highlight autonomy and privacy as central concerns. Many also raised broader ethical concerns around privacy and data misuse (44 participants) and accuracy in interpreting emotional or cognitive states (58 participants). We next explore ethics further.

### 4.3 Stage 3: Students' ethical concerns

We now present findings relating to RQ2 (*How do students prioritise ethical concerns regarding AI sensing and intervention systems?*). Figure 2 plots the median Likert ratings and interquartile range for each statement of a possible ethical concern in Table 3. Overall, statements relating to Privacy (mean=4.14) and Autonomy (mean=4.10) garnered the highest level of concern from participants, while Fairness (mean=2.73) was rated the least concerning.

There were greater variations in Learning, Transparency, and Accuracy. For Learning, concerns diverged depending on item framing: students were less worried about missing out on support without AI (e.g., Learning_1 and Learning_2, which are risks of not having AI), but expressed stronger concern that AI may not respect their individual learning pace (Learning_3, which is about a risk when AI is used). This suggests that students feared negative impacts when control over learning was ceded to the system more than they valued the potential benefits of AI. A similar pattern appeared in Transparency and Accuracy. Items reflecting system-centred performance with more distant concerns, such as *understanding how the algorithm works* (Transparency_1) or *the system failing to recognise when help is needed* (Accuracy_2), drew relatively lower concern and more variation. In contrast, items with immediate personal consequences, such as *not being informed about how data is used* (Transparency_2), or *misinterpreting behaviour and emotions* (Accuracy_1), elicited consistently higher concern. These results together highlight that students' priorities focus less on the technical promises of AIEd and more on the direct risks to their autonomy, privacy, and learning experience.

To better understand students' ethical priorities, we analysed the results from 1980 (15*132) paired comparisons participants made between two of the six key ethical principles of Autonomy, Transparency, Privacy, Fairness, Accuracy, and Learning Beneficence (Table.4). We first calculated the total 'wins' for each concept across each forced pairwise choice, which indicates how often this concept was prioritised as more important than another construct. The results reveal that Autonomy (N=515) and Privacy (N=462) received the highest total wins, indicating that participants placed the most importance on them. Transparency (N=269), Learning Beneficence (N=261), and Accuracy (N=248) scored lower, with Fairness (N=240) last. These results are consistent with the descriptive results from students' ratings.

To further quantify participants' preferences regarding specific ethical concerns, we applied the Bradley-Terry Model [20], which estimates the probability that one concept is more concerning than the alternative option and thus more likely to be selected. The probability of 'wins' confirmed that Autonomy (0.39) and Privacy (0.27) were by far the most important concerns, while Transparency (0.09), Learning Beneficence (0.09), Accuracy (0.08), and Fairness (0.08) were consistently less prioritised.

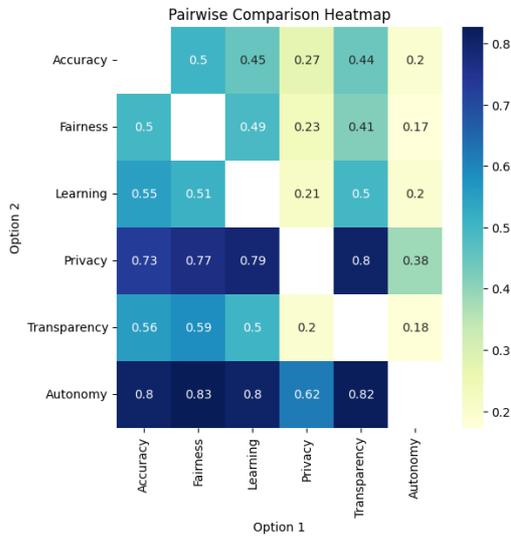

Figure 3: Ethical construct pairwise comparison heat map. Darker shades indicate a higher probability that the construct on the y-axis was selected as more concerning than the construct on the x-axis (See Table.4 for full ethical construct wording). Darker blue areas, such as those for autonomy and privacy, reflect stronger prioritisation of these concerns over others.

To visualise these preferences, Figure 3 presents a pairwise comparison heat map, showing the probability that one ethical construct was prioritised over another. The results confirm that Autonomy and Privacy were consistently judged as a higher ethical concern relative to Fairness, Transparency, Accuracy, and Learning Beneficence. It should be noted that even though Autonomy and Privacy are clearly more important to participants, the highest probability of wins in this context was 39%, indicating that no single concern completely overshadows the others within our participant sample.

### 4.4 Summary of Findings

Overall, the findings reveal that sensing causes strong negative affective responses and beliefs about learning experiences (H1 supported), with no significant differences between gaze-based attention detection and facial-based emotion detection (H2 rejected). Furthermore, the qualitative data consistently find discomfort, anxiety, and perceived disruption of learning caused by monitoring. Students expressed clear preferences for system-generated hints over human teacher assistance (H3 supported), largely due to a desire for agency and social concerns. Ethically speaking, students were most concerned about autonomy and privacy, suggesting that students highly value control over their data and learning process in AIEd interactions.

## 5 Discussion

We now discuss the significance of our study findings for AI sensing-intervention in terms of student affective responses, beliefs about learning impacts (5.1), and ethical concerns (5.2). Implications for the design and deployment of these new technologies follow (5.3).

### 5.1 Possibilities and challenges regarding AI sensing-intervention systems

Students recognised potential benefits of AI systems, especially if private and non-intrusive. This was most evident in their preference for system-generated hints over teacher-initiated assistance, reflected consistently in both quantitative and qualitative findings. In scenarios without intrusive sensing, students leaned towards agreeing that the system could provide personalised support and render learning easier. These results highlight the educational value of such AI.

Yet these perceived benefits proved fragile once students realised that interventions derived from sensing. Positive perceptions shifted to neutral or negative, with both emotion and attention detection provoking discomfort, anxiety, and beliefs that the system would be distracting. Thus, the very mechanism of sensing that promises personalised, responsive teaching also risks undermining students' learning experiences [53, 56]. AIEd design must therefore focus on preserving the benefits of intervention while avoiding intrusive sensing by exploring alternative sources of insight to guide learning support (see also Section 5.3).

Students also resisted teachers being summoned automatically, preferring control over when and how help was triggered. This reflected a strong need for learning autonomy [56]. Students also feared unintended social consequences, such as embarrassment when teachers are alerted. While sensing technology might enhance teacher awareness of students' learning [63, 139] and emotional status [42, 123], our participants framed such visibility as undermining privacy and autonomy. This again underscores the importance of situating AIEd within the classroom's social dynamics, not just its technical potential [54, 122].

Our results suggest that affective responses and learning beliefs tended to align. Positive emotions corresponded with more positive learning perceptions, and vice versa. This resonates with theories that learning is partly an emotional experience [114] and that students' feelings in learning processes strongly shape engagement and outcomes [78, 114]. AIEd designs that foster positive emotional experiences may therefore enhance students' motivation and learning [78].

### 5.2 Shape of students' ethical concerns

Students expressed clear ethical concerns about AI sensing-intervention systems. Ranked highest were the lack of consent before deployment and opt-out options (autonomy), followed by not being informed of data use and associated risks (one kind of transparency), and having learning struggles or emotions revealed to others (privacy). Concerns about erroneous interpretation of behaviour and emotions (accuracy) were moderate, while understanding how algorithms work (another kind of transparency) and fairness issues, such as algorithmic bias, evoked the least concern.

Pairwise comparisons reinforced this hierarchy, with students prioritising autonomy and privacy over potential learning benefits.

The prominence of consent and opt-out options reflect not only students' strong desire for greater agency in how technology is integrated into their learning, but also their foundational role: without being informed and able to choose, students cannot meaningfully evaluate or respond to other risks. Meanwhile, our measures captured only one aspect of autonomy – the ability to accept or decline system use. As both prior research [56] and our qualitative findings emphasise, autonomy in education also encompasses broader control over what and how one studies, and the ability to make learning choices without undue direction from technologies [56]. This aligns with higher education students' broader expectations of independence [134], and their preference for agency in AI-assisted classrooms [139]. Our participants echoed these wishes by suggesting improvements like making hints optional and allowing control over teacher requests.

Privacy concerns extended beyond data-centric issues [60] to social privacy. Students were particularly uneasy about learning struggles or emotional states being revealed to peers and teachers (Privacy_3), a concern rated even higher than data breach protection (Privacy_1). This resonates with prior HCI findings that technology use can create unexpected breaches of personal boundaries in everyday contexts [108, 120, 122], again highlighting the need to account for the social contexts of technology use [122].

Fairness and equal treatment received lower priority, which is surprising given well-documented risks of algorithmic bias (e.g., [11, 15, 41]) causing racial and gender discrimination (e.g., [23]). One potential explanation is students' limited awareness of this problem [110]. However, it is also possible that students were more focused on avoiding sensing elements altogether, prioritising opting out over equal access and treatment. Nonetheless, participants did raise inclusiveness concerns, particularly regarding neurodiversity. The fact that fairness was rated lower does not imply it was unimportant, but that other issues felt more pressing.

Taken together, students' ethical priorities around autonomy and privacy directly shape their negative responses to sensing and teacher-initiated interventions, and explain their preference for private, system-generated hints. From an HCI perspective, such ethical priorities can be operationalised through frameworks like value sensitive design [18] and participatory design [98], which embed learner values into technological decision-making. A value sensitive design approach to AIEd would accommodate ethical considerations, such as needs for autonomy and privacy, that could affect student well-being, trust, and learning.

## 5.3 Design Implications for AI sensing-intervention in education

We distil students' preferences and concerns into five design directions to guide the development of student-centred AIEd systems. The aim is to support student autonomy, privacy, trust, and well-being, while maintaining the pedagogical effectiveness of AIEd.

*1. Minimise intrusiveness in sensing.* Given students' aversion to both gaze-based attention and facial-based emotion detection, designers should prioritise alternative, less invasive ways to gauge learning progress. For instance, tracking time taken to complete tasks or responses to in-platform questions may be preferable. Assuming sensing technologies are used in learning and teaching, one-on-one individualised sensing deployed on personal devices may be preferable to classroom-wide sensing. Sensing on individual devices like laptops could allow students to opt in (e.g., choose whether to install or use the application) and configure how and when sensing is enabled. Such individual sensing would offer greater autonomy than indiscriminate classroom-wide sensing, which makes opting-out difficult.

*2. Refining system-delivered interventions.* Students generally favoured system-generated hints, but only when these preserved autonomy and comfort. Hints should therefore be student-triggered, lightweight, and customisable. Students wanted the option to decide when to receive support rather than having it imposed automatically. This suggests value in offering 'opt-in' prompts, allowing students to accept or dismiss the hints. Hints should be delivered unobtrusively, but targeted to aid progress without being overwhelming or distracting. Students should be able to adjust intervention frequency and style, thus retaining control. Finally, giving students transparency about why hints appear (e.g., "You've been on this step for a while, would you like a tip?") may help build trust and reduce feelings of surveillance [53].

*3. Redesigning teacher-mediated interventions.* Although students expressed discomfort with automatic teacher involvement, this intervention approach could be redesigned rather than discarded. First, students wanted control over when teachers were notified, a preference that runs counter to a current trend in AIEd where systems typically notify teachers directly (e.g., [31, 62]). Instead, systems could preserve and enhance student agency by enabling students to *request teacher help themselves* or to explicitly consent beforehand. Second, interventions could be less visible to peers. For example, a private digital chat or a subtle in-platform message rather than a teacher walking over in class seems desirable. Third, teachers could be provided with aggregated or anonymised signals (e.g., class-level trends of confusion) rather than individual flags, so they can adjust teaching without singling out students. Finally, when one-on-one support is necessary, teachers should be trained to frame interventions as collaborative offers of help rather than corrections, to reduce peer embarrassment.

*4. Dynamic consent and transparency.* Students sought agency not just at the start (opt-in/opt-out), but dynamically throughout use. Instead of a one-off consent form, systems could implement layered consent prompts at key moments, such as starting a difficult task (e.g., "Do you want sensing enabled during this task?", "Would you like your observed attentional status to be shared with your teacher?"). This dynamic consent approach respects students' varying comfort levels in different learning situations and reinforces trust in the system.

*5. Provide private feedback channels before involving teachers.* The system could first signal to students privately (e.g., "You may be struggling, would you like a tip?") *before* alerting teachers (e.g., "Would you like your teacher to help?"). This gives students a buffer stage to self-correct, accept a hint, or decide whether the teacher gets involved. This two-step approach could address both privacy and autonomy concerns while still enabling responsive interventions.

## 6 Limitations and future work

This study used video scenarios to elicit perceptions of AI sensing-intervention during puzzle tasks. While effective for capturing reactions, we did not fully simulate the experience of wrestling with a learning activity. Future work could employ actual systems to gather richer responses.

Our sample focused on Australian higher education students. Expectations, cultural norms, and ethical concerns can vary across countries, educational systems, and age groups. For instance, autonomy and privacy may be perceived differently in individualist versus collectivist cultures [57], and under different regulatory regimes (e.g., GDPR/AI Act in the EU). Broader studies, including those involving K-12 learners, could examine global and contextual variations and test the generalisability of the ethical prioritisation hierarchy.

Finally, our study reveals students' immediate reactions to hypothetical scenarios. The long-term effects of using AI systems in education, such as changes in behaviour or attitudes over time, remain unexplored. Future research might consider longitudinal studies to assess how students' experiences evolve with continued use. Nonetheless, the discomfort and ethical concerns we observed provide valuable guidance for developing technology that fosters students' positive attitudes and better aligns with their needs, values, and overall well-being in the long term.

## 7 Conclusion

This study investigated student perceptions and reactions to AI sensing-intervention used in higher education, focusing on affective responses, beliefs about learning impacts, and ethical concerns. While students valued the potential for personalised support, they consistently reacted negatively to gaze-based attention and facial-based emotion detection, citing anxiety, discomfort, and threats to autonomy and privacy. Preferences leaned toward system-delivered hints, provided these were private, lightweight, and under student control. Ethical priorities, especially autonomy and privacy, outweighed perceived learning benefits, with fairness, transparency, and accuracy considered secondary. This study contributes an empirical approach for examining student attitudes and value trade-offs in AIEd, offering guidance for creating systems that are both pedagogically effective and ethically robust. Ultimately, AIEd design must move beyond technical optimisation to embed agency, privacy, and social sensitivity at its core, ensuring technology serves education, not the other way around.

AI Sensing and Intervention in Higher Education: Student Perceptions of Learning Impacts, Affective Responses, and Ethical Priorities    CHI '26, April 13–17, 2026, Barcelona, Spain[74] Yelin Kim, Tolga Soyata and Reza Feyzi Behnagh. 2018. Towards Emotionally Aware Ai Smart Classroom: Current Issues and Directions for Engineering and Education. *IEEE Access* 6, 5308–5331.

[75] Abirami Kirubarajan, Dylan Young, Shawn Khan, Noelle Crasto, Mara Sobel and Dafna Sussman. 2022. Artificial Intelligence and Surgical Education: A Systematic Scoping Review of Interventions. *Journal of Surgical Education* 79, 2, 500–515.

[76] Ekaterina Kochmar, Dung Do Vu, Robert Belfer, Varun Gupta, Iulian Vlad Serban and Joelle Pineau. 2020. Automated Personalized Feedback Improves Learning Gains in an Intelligent Tutoring System. In *Proceedings of International conference on artificial intelligence in education*. Springer, 140–146.

[77] Kenneth R Koedinger, Emma Brunskill, Ryan SJd Baker, Elizabeth A McLaughlin and John Stamper. 2013. New Potentials for Data-Driven Intelligent Tutoring System Development and Optimization. *AI Magazine* 34, 3, 27–41.

[78] Jeya Amantha Kumar, Balakrishnan Muniandy and Wan Ahmad Jaafar Wan Yahaya. 2019. Exploring the Effects of Emotional Design and Emotional Intelligence in Multimedia-Based Learning: An Engineering Educational Perspective. *New Review of Hypermedia and Multimedia* 25, 1-2, 57–86.

[79] Priya C Kumar, Marshini Chetty, Tamara L Clegg and Jessica Vitak. 2019. Privacy and Security Considerations for Digital Technology Use in Elementary Schools. In *Proceedings of the 2019 CHI Conference on Human Factors in Computing Systems*. 1–13.

[80] Annabel Latham, Keeley Crockett, David McLean and Bruce Edmonds. 2012. A Conversational Intelligent Tutoring System to Automatically Predict Learning Styles. *Computers & Education* 59, 1, 95–109.

[81] Annabel Latham and Sean Goltz. 2019. A Survey of the General Public's Views on the Ethics of Using Ai in Education. In *Proceedings of Artificial Intelligence in Education: 20th International Conference, AIED 2019, Chicago, IL, USA, June 25-29, 2019, Proceedings, Part I 20*. Springer, 194–206.

[82] LuEttaMae Lawrence, Vanessa Echeverria, Kexin Yang, Vincent Aleven and Nikol Rummel. 2024. How Teachers Conceptualise Shared Control with an Ai Co-Orchestration Tool: A Multiyear Teacher-Centred Design Process. *British Journal of Educational Technology* 55, 3, 823–844.

[83] Xiaotian Vivian Li, Mary Beth Rosson and Jenay Robert. 2022. A Scenario-Based Exploration of Expected Usefulness, Privacy Concerns, and Adoption Likelihood of Learning Analytics. In *Proceedings of the Ninth ACM Conference on Learning@ Scale*. 48–59.

[84] Hao-Chiang Koong Lin, Chih-Hung Wu and Ya-Ping Hsueh. 2014. The Influence of Using Affective Tutoring System in Accounting Remedial Instruction on Learning Performance and Usability. *Computers in Human Behavior* 41, 514–522.

[85] Adam Linson, Yucheng Xu, Andrea R English and Robert B Fisher. 2022. Identifying Student Struggle by Analyzing Facial Movement during Asynchronous Video Lecture Viewing: Towards an Automated Tool to Support Instructors. In *Proceedings of International Conference on Artificial Intelligence in Education*. Springer, 53–65.

[86] Alex Jiahong Lu, Tawanna R Dillahunt, Gabriela Marcu and Mark S Ackerman. 2021. Data Work in Education: Enacting and Negotiating Care and Control in Teachers' Use of Data-Driven Classroom Surveillance Technology. *Proceedings of the ACM on Human-Computer Interaction* 5, CSCW2, 1–26.

[87] Alex Jiahong Lu, Gabriela Marcu, Mark S. Ackerman and Tawanna R Dillahunt. 2021. Coding Bias in the Use of Behavior Management Technologies: Uncovering Socio-Technical Consequences of Data-Driven Surveillance in Classrooms. In *Proceedings of Designing Interactive Systems Conference 2021*. Association for Computing Machinery, Virtual Event, USA, 508–522. http://dx.doi.org/10.1145/3461778.3462204

[88] Rose Luckin, Wayne Holmes, Mark Griffiths and Laurie B Forcier. 2016. Intelligence Unleashed: An Argument for Ai in Education.

[89] David Lundie. 2016. Authority, Autonomy and Automation: The Irreducibility of Pedagogy to Information Transactions. *Studies in philosophy and education* 35, 3, 279–291.

[90] Qianru Lyu, Wenli Chen, Junzhu Su, Kok Hui John Gerard Heng and Shuai Liu. 2023. How Peers Communicate without Words-an Exploratory Study of Hand Movements in Collaborative Learning Using Computer-Vision-Based Body Recognition Techniques. In *Proceedings of Artificial Intelligence in Education*. Springer Nature Switzerland, 316–326.

[91] Roberto Martinez-Maldonado, Andrew Clayphan and Judy Kay. 2015. Deploying and Visualising Teacher's Scripts of Small Group Activities in a Multi-Surface Classroom Ecology: A Study in-the-Wild. *Computer Supported Cooperative Work (CSCW)* 24, 2, 177–221. http://dx.doi.org/10.1007/s10606-015-9217-6

[92] Roberto Martinez-Maldonado, Vanessa Echeverria, Gloria Fernandez-Nieto, Lixiang Yan, Linxuan Zhao, Riordan Alfredo, Xinyu Li, Samantha Dix, Hollie Jaggard and Rosie Wotherspoon. 2023. Lessons Learnt from a Multimodal Learning Analytics Deployment in-the-Wild. *ACM Transactions on Computer-Human Interaction* 31, 1, 1–41.

[93] Bahar Memarian and Tenzin Doleck. 2023. Fairness, Accountability, Transparency, and Ethics (Fate) in Artificial Intelligence (Ai), and Higher Education: A Systematic Review. *Computers and Education: Artificial Intelligence*, 100152.

[94] Danaë Metaxa-Kakavouli, Kelly Wang, James A Landay and Jeff Hancock. 2018. Gender-Inclusive Design: Sense of Belonging and Bias in Web Interfaces. In *Proceedings of the 2018 CHI Conference on human factors in computing systems*. 1–6.

[95] Fengchun Miao, Wayne Holmes, Ronghuai Huang and Hui Zhang. 2021. *Ai and Education: A Guidance for Policymakers*. UNESCO Publishing,

[96] Jessica Morley, Luciano Floridi, Libby Kinsey and Anat Elhalal. 2020. From What to How: An Initial Review of Publicly Available Ai Ethics Tools, Methods and Research to Translate Principles into Practices. *Science and engineering ethics* 26, 4, 2141–2168.

[97] Maria Moundridou and Maria Virvou. 2002. Evaluating the Persona Effect of an Interface Agent in a Tutoring System. *Journal of computer assisted learning* 18, 3, 253–261.

[98] Michael J Muller and Sarah Kuhn. 1993. Participatory Design. *Communications of the ACM* 36, 6, 24–28.

[99] Chantal Mutimukwe, Olga Viberg, Lena-Maria Oberg and Teresa Cerratto-Pargman. 2022. Students' Privacy Concerns in Learning Analytics: Model Development. *British Journal of Educational Technology*.

[100] Andy Nguyen, Ha Ngan Ngo, Yvonne Hong, Belle Dang and Bich-Phuong Thi Nguyen. 2023. Ethical Principles for Artificial Intelligence in Education. *Education and Information Technologies* 28, 4, 4221–4241.

[101] Giovanna Nunes Vilaza, Kevin Doherty, Darragh McCashin, David Coyle, Jakob Bardram and Marguerite Barry. 2022. A Scoping Review of Ethics across Sigchi. In *Proceedings of the 2022 ACM Designing Interactive Systems Conference*. 137–154.

[102] Jum C Nunnally. 1978. Psychometric Theory (2nd Ed.).

[103] Amy Ogan. 2019. Reframing Classroom Sensing: Promise and Peril. *Interactions* 26, 6, 26–32.

[104] Prasoon Patidar, Tricia J Ngoon, John Zimmerman, Amy Ogan and Yuvraj Agarwal. 2024. Classid: Enabling Student Behavior Attribution from Ambient Classroom Sensing Systems. *Proceedings of the ACM on Interactive, Mobile, Wearable and Ubiquitous Technologies* 8, 2, 1–28.

[105] Yanyi Peng, Masato Kikuchi and Tadachika Ozono. 2023. Development and Experiment of Classroom Engagement Evaluation Mechanism during Real-Time Online Courses. In *Proceedings of International Conference on Artificial Intelligence in Education*. Springer, 590–601.

[106] Filipe Dwan Pereira, Samuel C Fonseca, Elaine HT Oliveira, Alexandra I Cristea, Henrik Bellhäuser, Luiz Rodrigues, David BF Oliveira, Seiji Isotani and Leandro SG Carvalho. 2021. Explaining Individual and Collective Programming Students' Behavior by Interpreting a Black-Box Predictive Model. *IEEE Access* 9, 117097–117119.

[107] Michael I Posner and Mary K Rothbart. 2007. Research on Attention Networks as a Model for the Integration of Psychological Science. *Annu. Rev. Psychol.* 58, 1, 1–23.

[108] Jenny Preece, Yvonne Rogers and Helen Sharp. 2002. *Interaction Design: Beyond Human-Computer Interaction*. John Wiley,

[109] Paul Prinsloo and Sharon Slade. 2016. Student Vulnerability, Agency and Learning Analytics: An Exploration. *Journal of Learning Analytics* 3, 1, 159–182. http://dx.doi.org/10.18608/jla.2016.31.10

[110] Shalini Ramachandran, Steven Matthew Cutchin and Sheree Fu. 2021. Raising Algorithm Bias Awareness among Computer Science Students through Library and Computer Science Instruction.

[111] Lee Richstone, Michael J Schwartz, Casey Seideman, Jeffrey Cadeddu, Sandra Marshall and Louis R Kavoussi. 2010. Eye Metrics as an Objective Assessment of Surgical Skill. *Annals of surgery* 252, 1, 177–182.

[112] Samantha Robertson, Tonya Nguyen and Niloufar Salehi. 2021. Modeling Assumptions Clash with the Real World: Transparency, Equity, and Community Challenges for Student Assignment Algorithms. In *Proceedings of the 2021 CHI Conference on Human Factors in Computing Systems*. 1–14.

[113] Paul Rodway and Astrid Schepman. 2023. The Impact of Adopting Ai Educational Technologies on Projected Course Satisfaction in University Students. *Computers and Education: Artificial Intelligence* 5, 100150.

[114] Jerry Rosiek. 2003. Emotional Scaffolding: An Exploration of the Teacher Knowledge at the Intersection of Student Emotion and the Subject Matter. *Journal of Teacher Education* 54, 5, 399–412.

[115] Xingran Ruan, Charaka Palansuriya and Aurora Constantin. 2023. Affective Dynamic Based Technique for Facial Emotion Recognition (Fer) to Support Intelligent Tutors in Education. In *Proceedings of International Conference on Artificial Intelligence in Education*. Springer, 774–779.

[116] Johnny Saldaña. 2021. The Coding Manual for Qualitative Researchers.

[117] Clara Schumacher and Dirk Ifenthaler. 2018. Features Students Really Expect from Learning Analytics. *Computers in human behavior* 78, 397–407.

[118] Donghee Shin. 2020. User Perceptions of Algorithmic Decisions in the Personalized Ai System: Perceptual Evaluation of Fairness, Accountability, Transparency, and Explainability. *Journal of Broadcasting & Electronic Media* 64, 4, 541–565.

[119] Donghee Shin and Yong Jin Park. 2019. Role of Fairness, Accountability, and Transparency in Algorithmic Affordance. *Computers in Human Behavior* 98, 277–284.

CHI '26, April 13–17, 2026, Barcelona, Spain  Bingyi Han et al.

## Appendix A Experimental Scenarios, Visual Materials and Text Instructions

We set the scene for the context, telling participants the scenario is in the tutorial class and as students they are working on their own tasks (Fig. 4). *Background picture and puzzle pictures were generated by DALL-E.

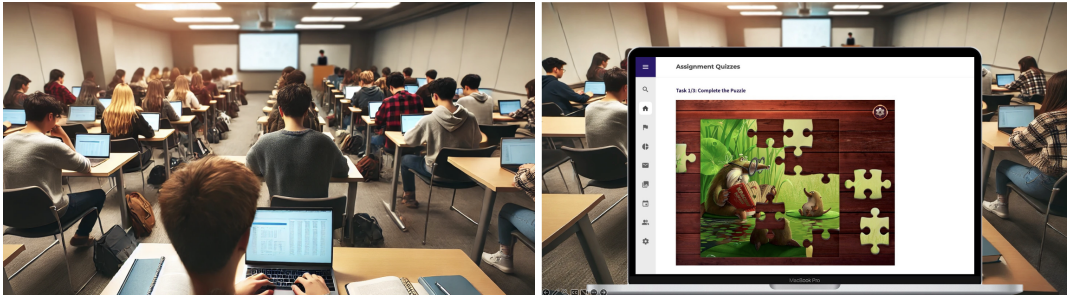

Figure 4: Visual Materials.

Scenario 1 *Sensing not in use + system hint (Fig.5)*: "Imagine that your university has introduced a new learning management system. You are in the tutorial class, practicing puzzle tasks. When the system thinks you are struggling, it sends you an automatic hint to help you."

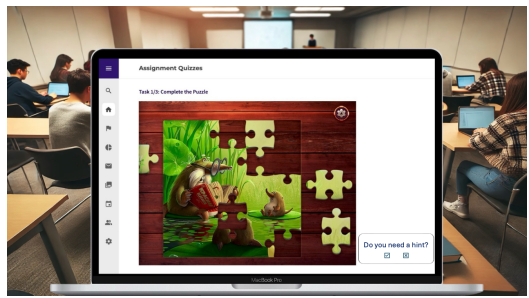

Figure 5: Visual Materials.

Scenario 2 *Sensing not in use + teacher assistance (Fig. 6)*: "Imagine that your university has introduced a new learning management system. You are in the tutorial class, practicing the puzzle tasks. When the system thinks you are struggling, it notifies your tutor. A notification then pops up, letting you know that your tutor is coming to help you."

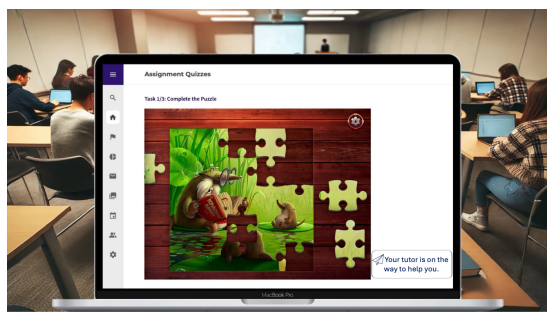

Figure 6: Visual Materials.

After completing the first two scenarios, we present the text: "In another system, the system observes your status to decide if you need help."

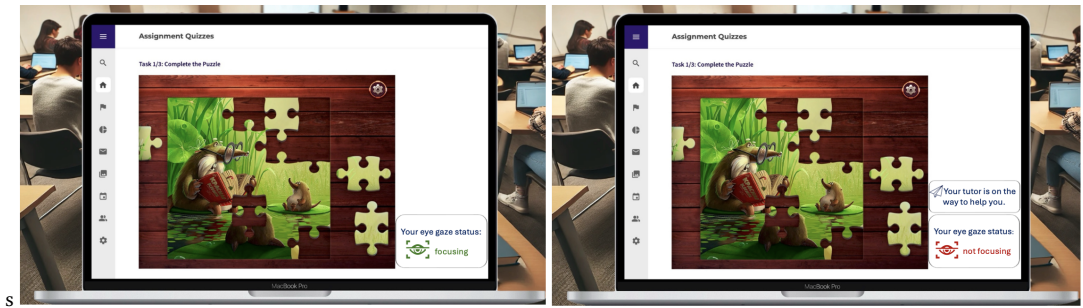

Figure 7: Visual Materials.

Scenario 3 *gaze-based attention + teacher assistance (Fig.7):* "You are in the tutorial class, practicing puzzle tasks. Your eye gaze is being tracked. When the system thinks you are struggling, it notifies your tutor. Your tutor will then come to help you."

Scenario 4 *gaze-based attention detection + system hint (Fig. 8):* "You are in the tutorial class, practicing puzzle tasks. Your eye gaze is being tracked. When the system thinks you are struggling, it sends you an automatic hint to help you."

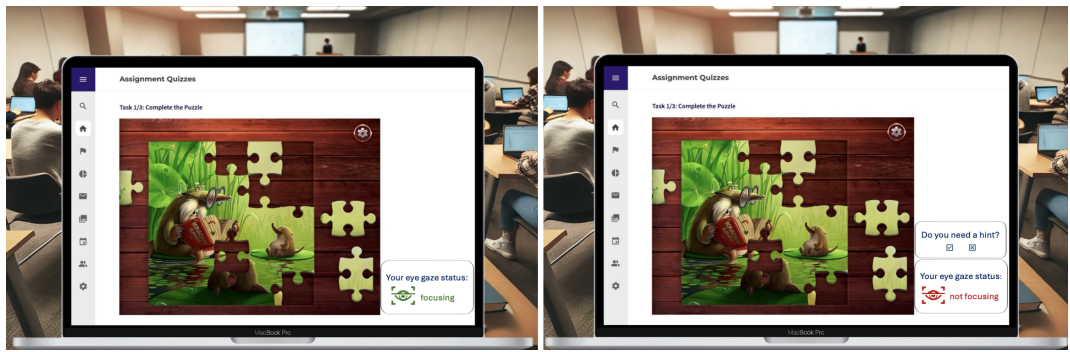

Figure 8: Visual Materials.

Scenario 5 *facial emotion-detection + teacher assistance (Fig. 9):* "You are in the tutorial class, practicing puzzle tasks. The camera on your laptop has read the emotions on your face. When the system thinks you are struggling, it notifies your tutor. Your tutor will then come to help you."

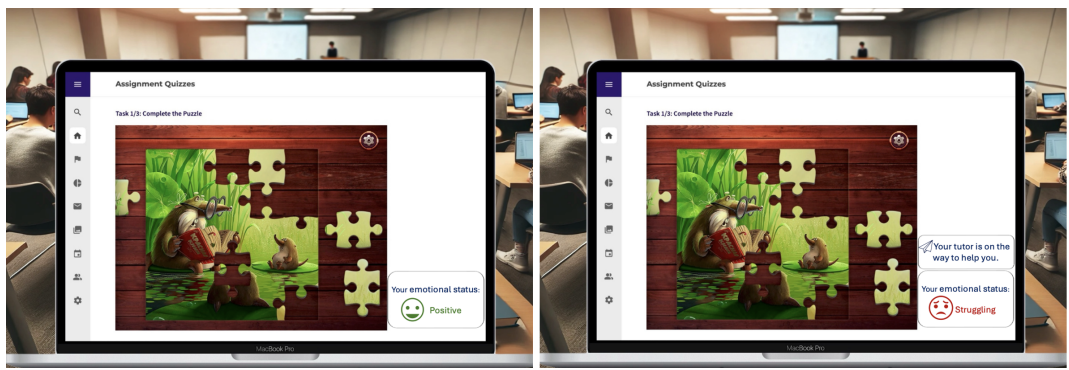

Figure 9: Visual Materials.

Scenario 6 *facial emotion-detection + system hint (Fig. 10):* "You are in the tutorial class, practicing puzzle tasks. The camera on your laptop has read the emotions on your face. When the system thinks you are struggling, it sends you an automatic hint to help you."

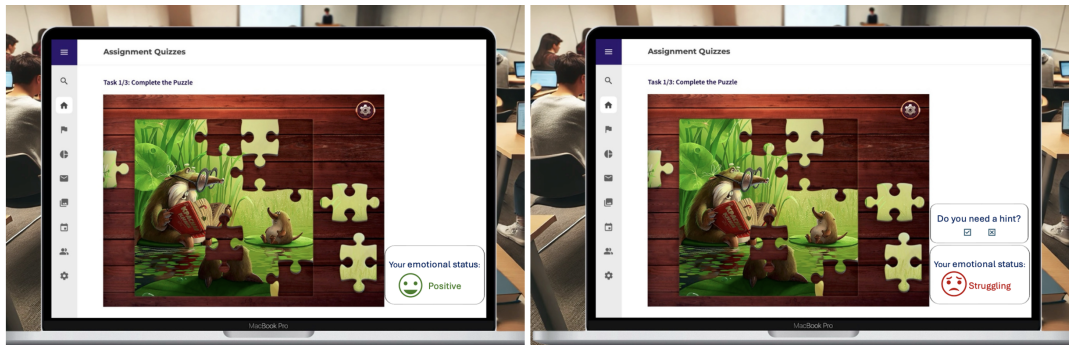

Figure 10: Visual Materials.

## Appendix B Synthesised view of intervention ratings (tables 7 and 8)

Table 9: Comparison of Ratings on the Intervention Form in Sensing Not-In-Use Scenario and Sensing In-Use Scenario

| Comparing results in 7 & Table 8 | Human Assist (No Sensing Scenario) Table 7 | | Human Assist (Sensing In-Use Scenario) Table 8 | | System Hint (No Sensing Scenario) Table 7 | | System Hint (Sensing In-Use Scenario) Table 8 | |
|---|---|---|---|---|---|---|---|---|
| | Mean | SD | Mean | SD | Mean | SD | Mean | SD |
| ***Belief about Learning*** | **3.09** | 0.92 | **2.29** | 0.86 | **3.47** | 0.82 | **2.73** | 0.97 |
| personalisation | **3.41** | 1.24 | **2.36** | 1.24 | **3.77** | 0.97 | **2.90** | 1.30 |
| efficiency | **3.29** | 1.25 | **2.23** | 1.20 | **3.91** | 0.95 | **2.88** | 1.31 |
| control of pace | **2.48** | 1.29 | **1.86** | 1.04 | **3.59** | 1.19 | **2.58** | 1.29 |
| grade performance | **3.44** | 1.19 | **2.49** | 1.18 | **3.56** | 0.96 | **2.84** | 1.15 |
| distraction | **3.17** | 1.19 | **4.05** | 1.08 | **2.77** | 1.13 | **3.74** | 1.15 |
| over-helping | **2.94** | 1.21 | **3.17** | 1.16 | **3.23** | 1.19 | **3.08** | 1.17 |
| ***Affective Responses*** | **2.26** | 0.95 | **1.70** | 0.88 | **3.47** | 0.94 | **2.21** | 0.95 |
| enjoyable | **2.62** | 2.00 | **1.73** | 1.02 | **3.52** | 1.06 | **2.27** | 1.17 |
| confident | **2.45** | 1.00 | **1.89** | 1.00 | **3.10** | 1.09 | **2.30** | 1.12 |
| anxious | **3.84** | 1.26 | **4.50** | 1.00 | **2.22** | 1.03 | **3.93** | 1.189 |
| uncomfortable | **3.96** | 1.18 | **4.55** | 0.93 | **2.56** | 1.23 | **4.11** | 1.138 |
| worried | **4.05** | 1.09 | **4.28** | 1.08 | **2.84** | 1.26 | **3.72** | 1.206 |
| fearing mistakes | **3.68** | 1.22 | **4.10** | 1.11 | **2.33** | 1.08 | **3.54** | 1.169 |

The response score is based on the Five-point Likert scale (1=Strongly disagree, 5=Strongly agree). Construct scores (shaded rows) were calculated by reverse-coding negatively worded items (e.g., 'distraction') so that higher values indicate more positive perceptions. Item-level scores are presented in their original direction.